\documentclass[journal=jctcce,manuscript=article,layout=traditional]{achemso}
\usepackage{chemformula} 
\usepackage[T1]{fontenc} 
\usepackage{caption}
\usepackage[utf8]{inputenc}
\usepackage{graphicx}
\usepackage{tabularx}
\usepackage{amsmath}
\usepackage{siunitx}
\usepackage{physics}
\usepackage{textgreek}
\usepackage{xfrac}
\usepackage{xcolor}
\usepackage{bm}
\usepackage{natbib}
\usepackage{bbold}
\usepackage{hyperref}
\usepackage{multicol}
\usepackage{appendix}
\usepackage[capitalise]{cleveref} 
\DeclareSIUnit\angstrom{\protect \text {Å}}

\def\agonem{$\mathrm{1A_g^-}$ }
\def\agtwom{$\mathrm{2A_g^-}$ }

\def\bup{$\mathrm{1B_u^+}$ }

\def\agone{$\mathrm{1A_g}$ }
\def\agtwo{$\mathrm{2A_g}$ }
\def\agthree{$\mathrm{3A_g}$ }
\def\buone{$\mathrm{1B_u}$ }
\def\butwo{$\mathrm{2B_u}$ }

\author{Timothy N. Georges}
\affiliation[University of Oxford]{Department of Chemistry, Physical and Theoretical Chemistry Laboratory,\\ University of Oxford, Oxford, OX1 3QZ, UK}
\alsoaffiliation[Balliol College]{Balliol College, University of Oxford, Oxford, OX1 3BJ, UK}
\email{timothy.georges@chem.ox.ac.uk}

\author{Darren J. Valentine}
\affiliation[University of Oxford]{Department of Chemistry, Physical and Theoretical Chemistry Laboratory,\\ University of Oxford, Oxford, OX1 3QZ, UK}
\alsoaffiliation[Balliol College]{Balliol College, University of Oxford, Oxford, OX1 3BJ, UK}

\author{William Barford}
\affiliation[University of Oxford]{Department of Chemistry, Physical and Theoretical Chemistry Laboratory,\\ University of Oxford, Oxford, OX1 3QZ, UK}
\alsoaffiliation[Balliol College]{Balliol College, University of Oxford, Oxford, OX1 3BJ, UK}
\email{william.barford@chem.ox.ac.uk}

\title[Nonadiabatic Quantum Dynamics of Hexatriene via the Extended Hubbard-Peierls Model and Time-Dependent DMRG]
  {Nonadiabatic Quantum Dynamics of Hexatriene via the Extended Hubbard-Peierls Model and Time-Dependent DMRG}

\begin{document}







\begin{abstract}
Propagating the electronic and nuclear degrees of freedom of a molecule simultaneously is a challenging task, because the number of vibronic states grows exponentially with system size. We use the adaptive time-dependent density matrix renormalization group to perform   simulations of the extended Hubbard-Peierls Hamiltonian, in which both the electronic and nuclear degrees of freedom are quantized.
We describe our approach, and demonstrate it through converged calculations of the nonadiabatic dynamics of hexatriene. 
This simulation predicts a lifetime of the `bright' $\mathrm{1B_u}$ electronic state of approximately\ 40 fs.
Our predictions for the population of the $\mathrm{1B_u}$ electronic state are in qualitative agreement with those obtained using the linear vibronic coupling model derived from the extended Hubbard-Peierls Hamiltonian in previous work.
\end{abstract}

\section{Introduction}\label{sec:intro}

Tensor network methods provide a way to calculate the dynamics of quantum systems exactly, up to some convergence parameters. In quantum chemistry, two widely used tensor network ans\"atze are tree tensor network states, used, for example, in the multilayer multiconfigurational time-dependent Hartree (ML-MCTDH) algorithm\cite{wangMultilayerFormulationMulticonfiguration2003,mantheMultilayerMulticonfigurationalTimedependent2008,vendrellMultilayerMulticonfigurationTimedependent2011}, and matrix product states, which can be propagated via a time-dependent density matrix renormalization group (TD-DMRG) algorithm, of which there exist several variations\cite{cazalillaTimeDependentDensityMatrixRenormalization2002,daleyTimedependentDensitymatrixRenormalizationgroup2004, whiteRealTimeEvolutionUsing2004,schollwockDensitymatrixRenormalizationGroup2005,schollwockDensitymatrixRenormalizationGroup2011,maTimedependentDensityMatrix2018,paeckelTimeevolutionMethodsMatrixproduct2019,xieTimedependentDensityMatrix2019,renTimeDependentDensity2022,xieTimeDependentDensityMatrix2026}.
These two methods are closely related\cite{larssonAtensornetworkview2024,larssonAccurateFullDimensionalComputations2026}, because matrix product states can be viewed as a special case of tree tensor network states and both ans\"atze can be propagated using projector-splitting integrators within the time-dependent variational principle\cite{lubichTimeIntegrationMulticonfiguration2015,haegemanUnifyingTimeEvolution2016,lindoyTimeEvolutionMLMCTDH2021}.
In the context of nonadiabatic dynamics, ML-MCTDH is commonly used to propagate vibronic coupling Hamiltonians\cite{mengMultilayerMCTDHStudy2013,lehrRoleVibronicCoupling2020,thalmannUnravelingChargeEnergy2026}, and it has also been used to propagate both electrons and nuclei quantum mechanically\cite{sasmalNonadiabaticQuantumDynamics2020}. TD-DMRG can be used to propagate vibronic coupling Hamiltonians\cite{baiardiLargeScaleQuantumDynamics2019,xieTimedependentDensityMatrix2019} and it has also been used to calculate the attosecond dynamics of the electronic structure Hamiltonian\cite{roncaTimeStepTargetingTimeDependent2017,baiardiElectronDynamicsTimeDependent2021,wahyutamaSimulatingRealTimeMolecular2024}.

In this work, we simulate the nonadiabatic dynamics of hexatriene using the extended Hubbard-Peierls Hamiltonian, defined in Section 2.  We  quantize both the electronic and nuclear degrees of freedom, and the system is propagated with adaptive TD-DMRG\cite{daleyTimedependentDensitymatrixRenormalizationgroup2004,whiteRealTimeEvolutionUsing2004}. 
Reduced-basis Hamiltonians are a well-established model for polyenes, where each site of the Hubbard Hamiltonian corresponds to a conjugated carbon $2p_z$-orbital\cite{schultenOriginLowlyingForbidden1972,tavanElectronicExcitationsFinite1987,schmidtElectronicExcitationsLong2012,BarfordBook,valentineHigherenergyTripletpairStates2020,karmakarLowlyingExcitedStates2022,manawaduPhotoexcitedStateDynamics2023,fabianPPPModelMinimum2026,nathElectronicStructureExcitedstate2026}. Using DMRG, there have been numerous \emph{static} calculations of quantum model Hamiltonians, such as the Holstein model\cite{zhangDensityMatrixApproach1998,tozerLocalizationLargePolarons2014}, the spin-Peierls model\cite{pearsonQuantizedLatticeDynamic2010}, the extended Hubbard-Peierls Hamiltonian,\cite{pearsonQuantizedLatticeDynamic2011,banerjeeGroundstateSpectralProperties2023a} and the Pariser-Parr-Pople Hamiltonian\cite{barfordBreakdownAdiabaticApproximation2002,valentineHigherenergyTripletpairStates2020}.

The \emph{dynamics} of the Holstein model (in which itinerant electrons are coupled  to a local Einstein oscillator via their charge density)\cite{liFiniteTemperatureTDDMRGCarrier2020,jansenFinitetemperatureDensitymatrixRenormalization2020,liGeneralChargeTransport2021,tenbrinkRealtimeNonadiabaticDynamics2022,geRolesNonlocalElectronphonon2024} and the Hubbard-Holstein model (in which itinerant, interacting electrons are coupled to a local Einstein oscillator via their charge density)\cite{noceraInterplayChargeSpin2014,tangTracesElectronPhononCoupling2022} have been extensively studied with TD-DMRG. Hashimoto and Ishihara used time-evolving block decimation (TEBD), an algorithm related to TD-DMRG, to propagate a spinless Holstein Hamiltonian supplemented with nearest-neighbor Coulomb interactions.\cite{hashimotoPhotoinducedChargeorderMelting2017} 
Mannouch and coworkers  used the Frenkel-Holstein Hamiltonian to model ultrafast relaxation of photoexcited states in $\pi$-conjugated polymers, also via TEBD.\cite{mannouch2018}
Menzler and coworkers propagated the spinless extended Hubbard-Peierls Hamiltonian using a mixed quantum-classical approach involving multi-trajectory Ehrenfest and time-evolving block decimation\cite{menzlerHybridQuantumclassicalMatrix2026}. Manawadu and coworkers explored the process of singlet fission in carotenoids by combining TD-DMRG evolution of the extended Hubbard-Peierls Hamiltonian with classical propagation of the nuclear degrees of freedom\cite{manawaduSingletTripletPairProduction2022,manawaduPhotoexcitedStateDynamics2023,manawaduDynamicalSimulationsCarotenoid2023}. However, the single-trajectory Ehrenfest algorithm used by Manawadu and coworkers fails to describe quantum nuclear effects such as wavepacket branching and zero-point energy. Arrachea and coworkers propagated the  quantum extended Hubbard-Peierls model using exact diagonalization, but due to the exponential scaling of the Hilbert space, they were limited to a model of butadiene with just two normal modes\cite{arracheaMicroscopicTheoryVibronic2003}.

By using adaptive TD-DMRG to propagate the fully quantum extended Hubbard-Peierls Hamiltonian, we have a viable method to simulate non-trivial molecules such as hexatriene\cite{daleyTimedependentDensitymatrixRenormalizationgroup2004,whiteRealTimeEvolutionUsing2004}.  As a consequence of our approach, we can describe the highly-correlated states of polyenes, while making no polynomial approximations to the potential energy surfaces. Moreover, unlike vibronic coupling models that include a small number of electronic states, our method includes coupling between all electronic states. 
However, since each electronic state now corresponds to a manifold of vibronic states that grows exponentially in system size, we cannot calculate the population of the entire manifold of states. To interpret the dynamics, we use the expectation value of the ionicity operator to derive a population estimator of the bright-state population.

In previous work, Georges and coworkers constructed a linear vibronic coupling (LVC) model of hexatriene from the same extended Hubbard-Peierls Hamiltonian used in this work\cite{georgesVibronicCouplingModel2026}. In this paper, we show that for up to $50~\unit{\fs}$ our simulations of the quantum extended Hubbard-Peierls Hamiltonian are in good agreement with this LVC model.

In the next section we introduce the extended Hubbard-Peierls model and the corresponding vibronic basis. Section 3 contains a discussion of the population estimator that is a proxy for the population of the `bright' \buone electronic state manifold, and there is a brief description of adaptive TD-DMRG in Section 4. Section 5 contains our results, and we conclude in Section 6. We also include two appendices. Appendix A describes how ionicity can be used to characterize the eigenstates of the extended Hubbard-Peierls model. Appendix B discusses TD-DMRG convergence;  we demonstrate convergence in the number of quantum oscillator levels per site, the number of block states, and the time step.

\section{Extended Hubbard-Peierls Hamiltonian}\label{sec:theory}

The extended Hubbard-Peierls (EHP) Hamiltonian is defined as:
\begin{equation}\label{eq:H_QUVP}
    \hat{H}_{\text{EHP}}=\hat{H}_{\text{EH}}+\hat{H}_{\text{P}}+\hat{H}_{\text{nuc}}+\hat{H}_{\epsilon}.
\end{equation}
$\hat{H}_{\text{EHP}}$ is the extended Hubbard Hamiltonian, $\hat{H}_{\text{EH}}$, supplemented with three terms: The  Peierls term that incorporates a linearized electron-nuclear coupling, $\hat{H}_{\text{P}}$; the nuclear kinetic and potential energy, $\hat{H}_{\text{nuc}}$; and an electron-hole symmetry breaking term, $\hat{H}_{\epsilon}$.

The extended Hubbard Hamiltonian is
\begin{equation}\label{eq:H_UVP_elec}
    \hat{H}_{\text{EH}}=U\sum_{n=1}^N\left(\hat{N}_{n\uparrow}-\frac{1}{2}\right)\left(\hat{N}_{n\downarrow}-\frac{1}{2}\right)
    +V\sum_{n=1}^{N-1}\left(\hat{N}_{n+1}-1\right)\left(\hat{N}_{n}-1\right)-2t_0\sum_{n=1}^{N-1}\hat{T}_n,
\end{equation}
where each site $n$  represents a carbon $2p_z$-orbital and $N$ is the number of carbon atoms. Since we consider a polyene at half-filling, the number of electrons is also equal to $N$.
$U$ and $V$ are the on-site and nearest-neighbor Coulomb interaction parameters. The final term on the right-hand side is the electron kinetic energy and $t_0$ is the average bond hopping integral. 

The number operator is
\begin{equation}
    \hat{N}_n=\sum_{\sigma}\hat{N}_{n\sigma}=\sum_\sigma \hat{c}_{n\sigma}^{\dagger}\hat{c}_{n\sigma}
\end{equation}
and the bond-order operator is
\begin{equation}
    \hat{T}_n=\frac{1}{2}\sum_{\sigma}\left(\hat{c}_{n\sigma}^{\dagger}\hat{c}_{n+1,\sigma}
    +\hat{c}_{n+1,\sigma}^{\dagger}\hat{c}_{n\sigma}\right).
\end{equation}
The creation (annihilation) operator $\hat{c}_{n\sigma}^{\dagger}$ ($\hat{c}_{n\sigma}$) creates (annihilates) an electron with spin $\sigma$ on site $n$.

The Peierls term couples variations in the electron kinetic energy term to changes in the bond length between neighboring C-atoms. It is defined as
\begin{equation}\label{ham:Peierls}
    \hat{H}_{\text{P}}=-2\sum_{n=1}^{N-1}(t_n-t_0)\hat{T}_n,
\end{equation}
where the transfer integrals are related to the nuclear displacements by
\begin{equation}
    t_n = t_0 - \alpha(x_{n+1}-x_n).
\end{equation}
$x_n$ is the nuclear displacement of the $n^{\text{th}}$  C-nucleus from its equilibrium position and $\alpha$ is the electron-nuclear coupling constant. To quantize the nuclear degrees of freedom, we express the position and momentum in terms of  local harmonic oscillator operators,
\begin{equation}\label{eq:position_as_qph_ops}
    x_n=\left(\frac{\hbar\omega_D}{4K}\right)^{\frac{1}{2}}\left(\hat{b}_n^{\dagger}+\hat{b}_n\right)
\end{equation}
and
\begin{equation}\label{eq:momentum_as_qph_ops}
    p_n=i\left(\frac{\hbar K}{\omega_D}\right)^{\frac{1}{2}}\left(\hat{b}_n^{\dagger}-\hat{b}_n\right).
\end{equation}
$\omega_D$ is the Debye frequency and $K$ is the force constant of a carbon-carbon $\sigma$ bond. The carbon mass is $m=2K/\omega_D^2$. The raising (lowering) operator $\hat{b}_n^{\dagger}$ ($\hat{b}_n$) raises (lowers) the quantum  level of the oscillator on site $n$, as defined by Eq.\ (\ref{Eq:16}). Substituting~\cref{eq:position_as_qph_ops,eq:momentum_as_qph_ops} into~\cref{ham:Peierls}, we have the quantized Peierls term,
\begin{equation}\label{eq:ham_quantum_peierls}
    \hat{H}_{\text{P}}=2\alpha\left(\frac{\hbar\omega_D}{2K}\right)^{\frac{1}{2}}\sum_{n=1}^{N-1}(\hat{B}_{n+1}-\hat{B}_n)\hat{T}_n,
\end{equation}
where $\hat{B}_n=x_n\sqrt{2K/\hbar\omega_D}$ is the dimensionless position operator of the nucleus on site $n$.

The classical nuclear kinetic and potential energy is
\begin{equation}\label{eq:ham_nuclear}
    \hat{H}_{\text{nuc}}=\sum_{n=1}^N\frac{p_n^2}{2m}+\frac{1}{2}K\sum_{n=1}^{N-1}\left(x_{n+1}-x_n\right)^2-2\alpha\Gamma\sum_{n=1}^{N-1}\left(x_{n+1}-x_n\right).
\end{equation}
The final term, parametrized by $\Gamma$, penalises deviations from a fixed chain length\cite{BarfordBook}. Substituting~\cref{eq:position_as_qph_ops,eq:momentum_as_qph_ops} into~\cref{eq:ham_nuclear} and taking into account open boundary conditions, we find
\begin{multline}\label{eq:ham_quantum_nuclear}
    \hat{H}_{\text{nuc}}=\hbar\omega_D\sum_{n=1}^N\left(\hat{b}_n^{\dagger}\hat{b}_n+\frac{1}{2}\right)
    -\frac{\hbar\omega_D}{2}\sum_{n=1}^{N-1}\hat{B}_{n+1}\hat{B}_n
    -2 \alpha\Gamma\left(\frac{\hbar\omega_D}{2K}\right)^{\frac{1}{2}}\left(\hat{B}_N-\hat{B}_1\right)\\
    -\frac{1}{4}\hbar\omega_D\left(\hat{b}_1^{\dagger}\hat{b}_1+\hat{b}_N^{\dagger}\hat{b}_N+1\right)
    +\frac{1}{8}\hbar\omega_D\left(\hat{b}_1^{\dagger 2}+  \hat{b}_1^2+\hat{b}_N^{\dagger 2}+\hat{b}_N^2\right).
\end{multline} 
The first term on the right-hand side of Eq.\ (\ref{eq:ham_quantum_nuclear}) is a sum over $N$ local (Einstein) harmonic oscillators, with a renormalized angular frequency, $\omega_D$. The second term is a nearest-neighbor coupling of these oscillators. Thus, the first two terms represent a quantized Debye phonon Hamiltonian, which can be diagonalized via a Bogoliubov transformation.\cite{Kittel} The normal modes are illustrated in Fig.\ 1 of Ref.\ \cite{georgesVibronicCouplingModel2026}.

The electronic extended Hubbard Hamiltonian, Eq.\ (\ref{eq:H_UVP_elec}), is invariant under the  electron-hole transformation,  $(\hat{N}_n - 1) \rightarrow -(\hat{N}_n - 1)$ or $\hat{c}_{n\sigma}^{\dagger} \rightarrow (-1)^n\hat{c}_{n\bar{\sigma}}$, where $\bar{\sigma}$ is the opposite spin-projection to ${\sigma}$.\cite{BarfordBook} The utility of electron-hole symmetry is that it distinguishes between covalent states, e.g., the `dark' $2^1A_g^-$ state, and ionic states, e.g., the `bright' $1^1B_u^+$ state. However, invariance to this symmetry also prevents internal conversion between the bright and dark state manifolds. To enable internal conversion between these manifolds, we supplement Eq.\ (\ref{eq:H_UVP_elec}) with a term that explicitly breaks electron-hole symmetry, namely
\begin{equation}
    \hat{H}_{\epsilon}=\sum_{n=1}^N\epsilon_n\left(\hat{N}_n-1\right),
\end{equation}
with $\epsilon_n$ a site-dependent potential energy listed in Table 1.

All but one of our model parameters are the same as those used by Georges and coworkers to construct a hexatriene linear vibronic coupling Hamiltonian from the extended Hubbard-Peierls Hamiltonian\cite{georgesVibronicCouplingModel2026}. These parameters are given in~\cref{tab:parameters} in Appendix~\ref{apx:characterising_vibronic_states}. The only different parameter  is $\Gamma$, which is determined by the constraint $\sum_{n=1}^{N-1}\left(x_{n+1}-x_n\right)=0$. Since quantum nuclear   fluctuations cause the ground state to be less dimerized than for classical nuclei\cite{barfordBreakdownAdiabaticApproximation2002}, a smaller value of $\Gamma=0.933$ is required to enforce a constant chain length. These parameters ensure an energy gap between the \buone and \agtwo excited states of $0.29~\unit{\eV}$ in the classical-nuclear ground state geometry that matches the vertical energy gap found by Komainda and coworkers with MSCASPT2\cite{komaindaInitioBenchmarkStudy2016}. Because dimerization is smaller for the quantum-nuclear ground state, however, the energy gap with a quantum-nuclear ground state is smaller, at $0.1~\unit{\eV}$. The properties of the electronic states of hexatriene are listed in~\cref{tab:hex_states,tab:hex_states_qph_gs} in Appendix~\ref{apx:characterising_vibronic_states}.

Our fully quantized EHP Hamiltonian lies beyond the Born-Oppenheimer approximation because of the simultaneous quantum treatment of electrons and nuclei. 
The full state vector consists of a linear combination of time-independent, many-particle vibronic basis states,
\begin{equation}
    \ket{\Psi(t)} = \sum_j c_j(t)\ket{\zeta_j}.
\end{equation}
Each of the many-particle vibronic basis states is in general a tensor product of the local site vibronic bases,
\begin{equation}
    \ket{\zeta_j}=\bigotimes_{n=1}^N|\sigma_n^{(j)}\rangle.
\end{equation}
The superscript $j$ indicates that the value of $\sigma_n$ is determined by the index $j$. The local vibronic states are a tensor product of local electronic and vibrational states,
\begin{equation}\label{eq:primitive_basis}
    |\sigma_n^{(j)}\rangle=|\phi_{\text{elec}}^{(j)}\rangle\otimes|v_{\text{nuc}}^{(j)}\rangle.
\end{equation}
For the  EHP Hamiltonian, there are four electronic states per site, $|\phi_{\text{elec}}^{(j)}\rangle\in\{\ket{0},\ket{\uparrow},\ket{\downarrow},\ket{\uparrow\downarrow}\}$. 

Although the number of electrons is a conserved quantity, the number of harmonic oscillator quanta per site is not. For a given vibronic basis, the parameter $D$ defines the maximum number of oscillator levels per site. Thus, the local  oscillator states are
\begin{equation}\label{Eq:16}
    |v_{\text{nuc}}^{(j)}\rangle = \frac{(\hat{b}_n^{\dagger})^{v}}{\sqrt{v !}}\ket{0}\rangle,
\end{equation}
where $v$ lies in the range $0\leq v\leq (D-1)$. $\ket{0}\rangle$ is the ground vibrational state of the $n^{\text{th}}$ local oscillator. In our TD-DMRG simulations, $D$  ranges from 2 to 6. Convergence with respect to $D$ is illustrated in Fig.\ 1.

Limiting the electronic states to the $S_z=0$ subspace, the Hilbert space size is thus
\begin{eqnarray}
\nonumber
    \text{Size}(\mathcal{H}_{\text{EHP}})= && \begin{pmatrix} N \\ N/2 \end{pmatrix}^2 \times D^N\\
    = && 400 \times D^6,
\end{eqnarray}
when $N = 6$.  When $D = 6$ the total fermionic-bosonic Hilbert space size  is 18,662,400. DMRG truncates this space to approximately\ $10^6$  states.

\section{Population Estimator}\label{sec:population_estimator}
Assuming $\mathrm{C_{2h}}$ geometry, the three lowest-lying adiabatic electronic states of the extended Hubbard-Peierls Hamiltonian are the $\mathrm{1A_g}$, \buone and \agtwo states. In the ground state geometry, the \agtwo state lies above the \buone state, but the \agtwo state falls below the \buone state due to the displacement along the optic mode\cite{georgesVibronicCouplingModel2026}. In the vibronic basis, each electronic state forms a manifold of $D^N$ vibronic states with alternating $\mathrm{A_g}$ and $\mathrm{B_u}$ symmetry. To find electronic state populations, it is too expensive to calculate the properties of all $D^N$ vibronic states included in the manifold of a single electronic state. We therefore need a population estimator to identify the population of the \buone electronic manifold.

In Appendix~\ref{apx:characterising_vibronic_states} we use the minimal case of $D=2$ (corresponding to a maximum of one excitation of each site quantum oscillator)  to show that the \agone vibronic manifold is well-separated in energy from the excited-state manifolds. Then, we show that the bright \buone electronic state can easily be identified from other low-lying dark excited states by its ionicity. Ionicity is a measure of how many sites are doubly occupied or unoccupied by electrons,
\begin{equation}
    \hat{I}=\sum_{n=1}^N\left(\hat{N}_n-1\right)^2.
\end{equation}
Although the \buone electronic state is ionic in character, the other low-lying singlet excitations have covalent character, allowing us to estimate the \buone population\cite{valentineHigherenergyTripletpairStates2020}.

The time-dependent expectation value of the ionicity is
\begin{equation}\label{Eq:19}
     I(t) = \sum_a |c_a(t)|^2 \langle a |\hat{I}|a \rangle + \sum_{a\ne b} c_a^*(t)c_b(t) \langle a |\hat{I}|b \rangle,
\end{equation}
where $c_a(t)$ is the  probability amplitude of the adiabatic electronic state $a$ in the evolving state vector, $|\Psi(t)\rangle$.
If now we assume that only the two lowest-lying electronic excited states are populated after photoexcitation
and if we neglect the cross-term in~\cref{Eq:19}, we have
\begin{equation}
    I(t) = I_{\mathrm{1B_u}}P_{\mathrm{1B_u}}(t) + I_{\mathrm{2A_g}}P_{\mathrm{2A_g}}(t),
\end{equation}
where $I_a = \langle a |\hat{I}|a \rangle$ is the known ionicity of the adiabatic state $a$ and $P_a(t)=|c_a(t)|^2$ is the electronic adiabatic state population to be determined. 
Since the sum of populations equals unity, we can determine the approximate \buone population by
\begin{equation}\label{eq:population_estimator}
    P_{\mathrm{1B_u}}(t) = \frac{I(t)-I_{\mathrm{2A_g}}}{I_{\mathrm{1B_u}}-I_{\mathrm{2A_g}}},
\end{equation}
where $I_{\mathrm{1B_u}}=3.15$ and $I_{\mathrm{2A_g}}=1.91$ are found in the Born-Oppenheimer ground state geometry. 

We note that Eq.\ (\ref{eq:population_estimator})  is an approximation for the \buone population. This is because, first, as already noted, the cross-term is neglected in~\cref{Eq:19}. Second, unlike other common nonadiabatic dynamics methods, we cannot  start the simulation in the vertical \buone state, as the \buone electronic state becomes a manifold of $D^N$ vibronic states in our basis.
Thus, our initial state, $\ket{\Psi(0)}$, is the transition dipole moment operator applied to the vibronic ground state, $\ket{\mathrm{GS}}$, i.e.,
\begin{equation}\label{eq:initial_state}
    \ket{\Psi(0)}=\hat{\mu}\ket{\mathrm{GS}},
\end{equation}
where the transition dipole moment operator is
\begin{equation}\label{eq:tdm_operator}
    \hat{\mu}=\sum_{n=1}^N\left(n-\frac{N+1}{2}\right)  \left(\hat{N}_n-1\right).
\end{equation}
Since $\hat{\mu}$ is odd under a two-fold rotation, while  $\ket{\mathrm{GS}}$ has $\mathrm{A_g}$ symmetry, $\ket{\Psi(0)}$ necessarily contains eigenstates of $\mathrm{B_u}$ symmetry that lie outside of the \buone electronic manifold.  
As a consequence of these approximations, $I(t=0)$ exceeds $I_{\mathrm{1B_u}}$, so that (as shown in Fig.\ 1), $P_{\mathrm{1B_u}}(t=0) \simeq 1.1$. Assuming that these errors remain constant during the dynamics, the error in our population estimator is ca.\ 10\%. 

\section{Adaptive Time-Dependent DMRG}\label{sec:DMRG}
There exist many formulations of time-dependent DMRG. Here, we use adaptive time-dependent DMRG (tDMRG), which was developed simultaneously by Daley and coworkers and White and Feiguin\cite{daleyTimedependentDensitymatrixRenormalizationgroup2004,whiteRealTimeEvolutionUsing2004}. The algorithmic details of the tDMRG implementation used in this work are described by Manawadu and coworkers\cite{manawaduDynamicalSimulationsCarotenoid2023}. The extension of this work is the addition of $D$ local harmonic oscillator basis states to each site, as described in~\cref{eq:primitive_basis}, and the addition of the quantum nuclear terms,~\cref{eq:ham_quantum_nuclear,eq:ham_quantum_peierls}, to the Hamiltonian.

We now summarize some of the key points of the tDMRG algorithm.\cite{manawaduDynamicalSimulationsCarotenoid2023} In this formulation, the Hamiltonian is expressed as a sum of local-bond Hamiltonians,
\begin{equation}
    \hat{H}_{\text{EHP}}=\sum_{n=1}^{N-1}\hat{H}_{n,n+1}.
\end{equation}
Each local Hamiltonian $\hat{H}_{n,n+1}$ only contains terms that act on or between sites $n$ and $n+1$. Therefore, this approach can only be used on Hamiltonians with at most nearest-neighbor connectivity.

The time evolution operator is found approximately with the second order Suzuki-Trotter decomposition,
\begin{equation}
    \mathrm{e}^{-i\hat{H}_{\text{EHP}}\delta t/\hbar}\approx
    \mathrm{e}^{-i\hat{H}_{1,2}\delta t/2\hbar}\mathrm{e}^{-i\hat{H}_{2,3}\delta t/2\hbar}\cdots
    \mathrm{e}^{-i\hat{H}_{N-1,N}\delta t/\hbar}\cdots
    \mathrm{e}^{-i\hat{H}_{2,3}\delta t/2\hbar}\mathrm{e}^{-i\hat{H}_{1,2}\delta t/2\hbar}
    +\mathcal{O}(\delta t^3).
\end{equation}
The full time evolution operator is approximated as a product of local bond evolution operators, $\mathrm{e}^{-i\hat{H}_{i,i+1}\delta t/2\hbar}$. Using the Baker-Campbell-Hausdorff theorem, it can be shown that the error is order $\mathcal{O}(\delta t^3)$. Evolving the state for a time $T$, $T/\delta t$ time steps are required, so the overall error of tDMRG is $\mathcal{O}(\delta t^2)$.

At each step in the tDMRG sweep, the local-bond evolution operator is applied exactly between all single-site blocks. This is done by diagonalizing the local-bond Hamiltonians before dynamics and rotating into this eigenstate basis to exactly perform time evolution of the local-bond Hamiltonian. Blocks $2$ and $3$ are always single-site blocks, but at the start or end of a sweep, either of the  terminal blocks $1$ and $4$ will also consist of a single site.

In our DMRG algorithm, the total spin projection, $S_z$, and the number of electrons are explicitly conserved quantities. Since $[ \hat{H}_{\text{EHP}},\hat{S}^2] = 0$, the total spin is also conserved.

We investigate the nonadiabatic dynamics of hexatriene for $100~\unit{\fs}$ after photoexcitation. We keep $m=800$ states in the reduced density matrix, use a time step of $0.003~\unit{\fs}$ and measure observables every $0.3~\unit{\fs}$. The reduced density matrix is composed of $50\%$ the evolving state, and $25\%$ of the ground state and the initial state.

tDMRG simulations are preceded by infinite and finite DMRG calculations which calculate the ground state of the EHP Hamiltonian\cite{whiteDensitymatrixAlgorithmsQuantum1993,manawaduDynamicalSimulationsCarotenoid2023}. 
To simulate photoexcitation, our initial state is defined by Eq.\ (\ref{eq:initial_state}).

\section{Results}\label{sec:results}
\begin{figure}
    \centering
   \includegraphics[width=0.5\linewidth]{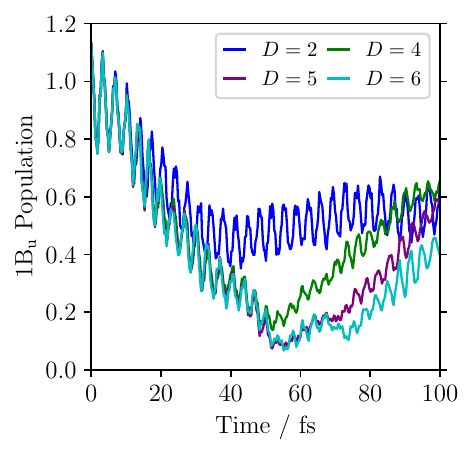}
    \caption{The population of the \buone electronic manifold found with our ionicity population estimator, plotted against time, varying the number of quantum oscillator levels per site, $D$. The $D=5$ result is converged to the $D=6$ result up to $70~\unit{\fs}$. There is almost complete internal conversion after $52~\unit{\fs}$, after which the \buone population increases again. }
    \label{fig:hexatriene_dynamics_and_D_conv}
\end{figure}

The \buone population is plotted against time in~\cref{fig:hexatriene_dynamics_and_D_conv}. Here, \buone population means the population of the vibronic manifold associated with the adiabatic, electronic \buone state, and it is found from the total ionicity using the population estimator given in~\cref{eq:population_estimator}. We first notice, as discussed in Section 3, that the calculated population of the \buone manifold exceeds unity  by ca.\ 10\% at very early times. 

To establish convergence with respect to  the number of quantum oscillator levels per site, $D$,  we increase $D$ from $2$ to $6$. Keeping just $D=2$ levels per site leads to very inaccurate dynamics after $20~\unit{\fs}$. Our best simulation with $D=6$  converges to the $D=5$ result until $70~\unit{\fs}$, so this is the time up to which we consider our $D=6$ result fully converged. The convergence in the other sources of error in the tDMRG, $m$ and $\delta t$, is shown in Appendix~\ref{apx:convergence}.

Considering the $D=6$ result in~\cref{fig:hexatriene_dynamics_and_D_conv}, we predict almost complete internal conversion from the \buone to the \agtwo electronic manifolds in $52~\unit{\fs}$. After reaching a minimum of $0.07$, the \buone population increases for the remainder of the dynamics. Up to $50~\unit{\fs}$, there are regular oscillations in the \buone population estimator. These have a period of $3.33~\unit{\fs}$, which corresponds to an energy gap of $1.24~\unit{\eV}$. We assign these oscillations to the population of other  excited electronic states in the initial state $\ket{\Psi(0)}$, as discussed in Section 3.

\begin{figure}
    \centering
   \includegraphics[width=0.5\linewidth]{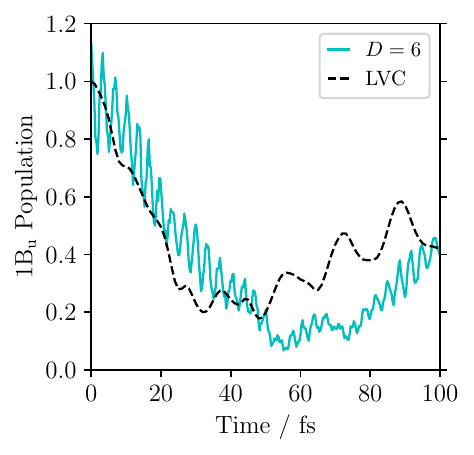}
    \caption{The population of the \buone electronic manifold plotted against time, for two different methods. This work: tDMRG propagation of the EHP Hamiltonian, using an ionicity population estimator for the \buone population and $D=6$ oscillator levels per site. Previous work: Diabatic \buone populations from the SILP propagation of a linear vibronic coupling (LVC) Hamiltonian constructed from the extended Hubbard-Peierls Hamiltonian\cite{georgesVibronicCouplingModel2026}.}
    \label{fig:lvc_hexatriene_dynamics}
\end{figure}

\cref{fig:lvc_hexatriene_dynamics} shows our $D=6$ result compared to the previous work of Georges and coworkers, who constructed a linear vibronic coupling (LVC) Hamiltonian from the extended Hubbard-Peierls Hamiltonian and propagated it exactly with the short iterative Lanczos propagator method (SILP)\cite{georgesVibronicCouplingModel2026}. Up to $50~\unit{\fs}$, there is qualitative agreement between their \buone diabatic populations and the \buone population derived from our population estimator. The initial rate of decay is similar, but the LVC model predicts a smaller amount of internal conversion, where $P_{\mathrm{1B_u}}=0.2$, reached at an earlier time of $32~\unit{\fs}$. The LVC model has not only a different population definition, but also a different initial state; the LVC dynamics begins in the \buone diabatic state in the ground vibrational wavefunction. Therefore, it lacks any population of higher-energy bright states along with the accompanying oscillations seen in the tDMRG result.

\section{Conclusions}\label{sec:conclusion}
In this paper, we describe a method to simulate the nonadiabatic dynamics of photoexcited polyene molecules fully quantum-mechanically. To reduce computational complexity, we use the extended Hubbard-Peierls Hamiltonian, rather than \textit{ab initio} approaches. The nuclear degrees of freedom are represented by coupled on-site quantum oscillators.  We propagate the electronic and nuclear degrees of freedom with adaptive time-dependent DMRG.

For hexatriene, we demonstrate convergence in the number of oscillator levels, DMRG system-block size, and the time step. Due to the computational cost of these calculations, we are unable to target longer polyenes. The initial state is the dipole moment operator acting on the ground state and we use the computed ionicity as a proxy to approximate the electronic state populations. We show almost complete internal conversion from the \buone to the \agtwo electronic manifolds in $52~\unit{\fs}$. Up to $50~\unit{\fs}$, there is qualitative agreement with the \buone populations found from a linear vibronic coupling model of hexatriene derived from the  extended Hubbard-Peierls Hamiltonian in previous work by Georges and coworkers\cite{georgesVibronicCouplingModel2026}.

Future work should focus on reducing the cost of simulations, possibly via local basis optimization\cite{zhangDensityMatrixApproach1998, brocktMatrixproductstateMethodDynamical2015,tenbrinkRealtimeNonadiabaticDynamics2022}. If larger simulations become achievable, fully quantum, dynamical simulations of the  extended Hubbard-Peierls  Hamiltonian could also be used to describe carotenoids and other conjugated polymers.

\begin{acknowledgement}
T.N.G. is grateful for financial support from the Department of Chemistry, University of Oxford, the Engineering and Physical Sciences Research Council (grant number EP/W524311/1) and Balliol College, University of Oxford for a Balliol Scholarship and Jowett Scholarship. T.N.G. thanks Dilhan Manawadu for useful discussions regarding DMRG. D.J.V. would like to thank the EPSRC Centre for Doctoral Training, Theory and Modelling in Chemical Sciences (grant number EP/L015722/1) for financial support and Balliol College, University of Oxford for the Foley-B\'ejar Scholarship. The authors would like to acknowledge the use of the University of Oxford Advanced Research Computing (ARC) facility in carrying out this work.
\end{acknowledgement}



\bibliography{main.bib}

\newpage
\appendixpage
\appendix
\section{Characterising Vibronic States}\label{apx:characterising_vibronic_states}
\subsection{Parameters}
The parameters used in this work are, with the exception of $\Gamma$, the same as those used in previous work by Georges and coworkers\cite{georgesVibronicCouplingModel2026}. These parameters are given in~\cref{tab:parameters}.

\begin{table}[ht!]
    \centering
    \begin{tabular}{|c|c|c|}
        \hline
        \textbf{Parameter} & \textbf{Symbol} & \textbf{Value} \\
        \hline
        On-site Coulomb & $U$ & $7.25~\unit{\eV}$  \\
        \hline
        Nearest-neighbor Coulomb & $V$ & $3.35~\unit{\eV}$  \\
        \hline
        Bond-hopping integral & $t_0$ & $2.40~\unit{\eV}$  \\
        \hline
        $\sigma$-bond force constant & $K$ & $46.0~\unit{\eV\angstrom^{-2}}$  \\
        \hline
        Debye energy & $\hbar\omega_D$ & $0.20~\unit{\eV}$  \\
        \hline
        Electron-nuclear coupling & $\alpha$ & $4.593~\unit{\eV\angstrom^{-1}}$  \\
        \hline
        Constant chain length & $\Gamma$ & $0.933$  \\
        \hline
        Symmetry breaking term & $\epsilon_1$ & $1.702~\unit{\eV}$  \\
        \hline
        Symmetry breaking term & $\epsilon_2$ & $0.473~\unit{\eV}$  \\
        \hline
        Symmetry breaking term & $\epsilon_3$ & $-1.982~\unit{\eV}$  \\
        \hline
    \end{tabular}
    \caption{Parameters of the EHP Hamiltonian. All parameters, except $\Gamma$,  are taken from previous work on the hexatriene extended Hubbard-Peierls Hamiltonian by Georges and coworkers\cite{georgesVibronicCouplingModel2026}.}
    \label{tab:parameters}
\end{table}

\subsection{Electron-Hole Symmetric Vibronic States}
The electronic eigenstates of the extended Hubbard-Peierls Hamiltonian, with no electron-hole symmetry breaking term $\hat{H}_{\epsilon}$, will be referred to as the electron-hole symmetric (EHS) states. Expressed in terms of the $\mathrm{C_{2h}}$ point group term symbols, the four lowest-lying singlet states are $\mathrm{1A_g^-}$, $\mathrm{1B_u^+}$, $\mathrm{2A_g^-}$ and $\mathrm{1B_u^-}$. Solving the EHP Hamiltonian for the minimal case of $N=6$ and $D=2$, we can calculate the properties of the first $200$ eigenstates. Since there are $D^N= 2^6 = 64$ vibronic states corresponding to the vibronic manifold of each electronic state, we expect these $200$ states to include the $\mathrm{1A_g^-}$, $\mathrm{1B_u^+}$ and $\mathrm{2A_g^-}$ electronic manifolds. We start our description of vibronic states with the EHS states, because their symmetry makes their interpretation easier.

\begin{figure}
    \centering
    \includegraphics[width=0.5\linewidth]{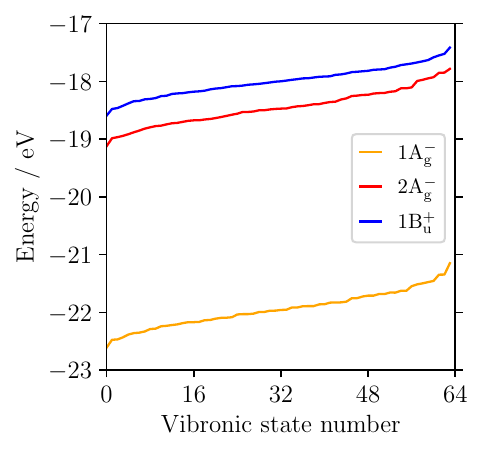}
    \caption{Energies of the first $200$ vibronic states for hexatriene with the EHP Hamiltonian. There is no symmetry breaking term and $D=2$. The positive and negative electron-hole symmetry sectors are separated, allowing the \agtwom and \bup electronic manifolds to be identified.}
    \label{fig:qph2_phs1_energies}
\end{figure}

\cref{fig:qph2_phs1_energies} shows the energies of the first $200$ vibronic states for the electron-hole symmetric EHP Hamiltonian. Since we can specify electron-hole symmetry, the \bup and \agtwom sectors can be separated, despite the fact they overlap in energy. The \agonem sector is well separated in energy.

\begin{figure}
    \centering
    \includegraphics[width=0.5\linewidth]{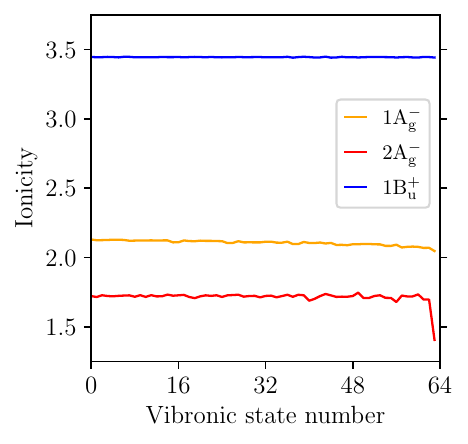}
    \caption{Ionicities of the first $200$ vibronic states for hexatriene with the EHP Hamiltonian. There is no symmetry breaking term and $D=2$. The bright \bup electronic manifold has much higher ionicity than the other low-lying states.}
    \label{fig:qph2_phs1_ionicity}
\end{figure}

\cref{fig:qph2_phs1_ionicity} shows the ionicities of the first $200$ vibronic states for the electron-hole symmetric EHP Hamiltonian. We can assign the first $64$ states of negative electron-hole symmetry to the \agonem state, because we know from the energy calculation that they are well-separated in energy. The \agtwom state has a lower ionicity than the ground state, whereas the \bup state has a much larger ionicity at $3.45$. This value is constant across the range of vibronic eigenstates, so it is a reliable measure of \bup character for whatever nuclear motion occurs during dynamics, assuming that only the \bup and \agtwom are populated.

\begin{figure}
    \centering
    \includegraphics[width=0.5\linewidth]{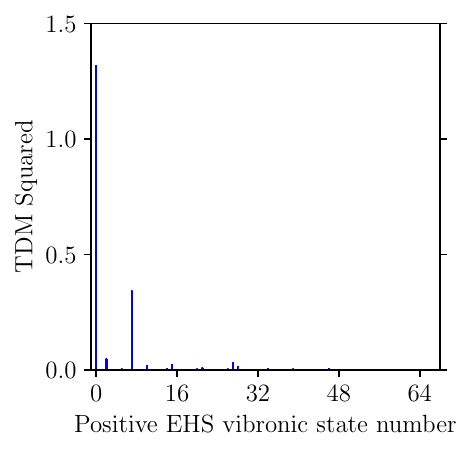}
    \caption{The square of the transition dipole moment, $|\mel{\mathrm{S^+_{\textit{j}}}}{\hat{\mu}}{\mathrm{GS}}|^2$, of the first $68$ vibronic states in the positive electron-hole symmetry (EHS) sector with the ground vibrational \agonem state. These results are for hexatriene with the EHP Hamiltonian without the symmetry breaking term and $D=2$.}
    \label{fig:qph2_phs1_tdm_with_1Ag0}
\end{figure}

Finally,~\cref{fig:qph2_phs1_tdm_with_1Ag0} shows the squared transition dipole moment, $|\mel{\mathrm{S^+_{\textit{j}}}}{\hat{\mu}}{\mathrm{GS}}|^2$, of the first $68$ positive electron-hole symmetry vibronic states, labelled $\mathrm{S^+_{\textit{j}}}$, with the ground vibrational \agonem state, $\ket{\mathrm{GS}}$. The state of greatest overlap with the ground state is the $0-0$ transition. There is not a large peak at the $64^{\text{th}}$ state, because the second electronic state manifold in the positive electron-hole symmetry sector is not dipole connected to the ground state.

\subsection{EHP Vibronic States}
\begin{figure}
    \centering
    \includegraphics[width=0.5\linewidth]{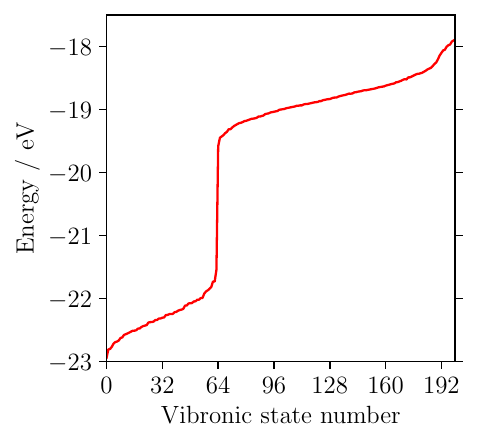}
    \caption{Energies of the first $200$ vibronic states for hexatriene with the EHP Hamiltonian and $D=2$. There is a clear energy separation between the manifold of the \agone electronic state, and the \agtwo and \buone electronic manifolds. However, the \agtwo and \buone states cannot be identified by their energies. }
    \label{fig:qph2_phs0_energies}
\end{figure}

With the electron-hole symmetry breaking term  (Eq.\ (12)) included in the EHP Hamiltonian, we can no longer specify the electron-hole symmetry of the eigenstates of the Hamiltonian. Therefore, the lowest-lying states are $\mathrm{1A_g}$, $\mathrm{1B_u}$, $\mathrm{2A_g}$ and $\mathrm{2B_u}$. The energies of the lowest-lying $200$ vibronic eigenstates are shown in~\cref{fig:qph2_phs0_energies}. As for the EHS states, the \agone ground state is well-separated in energy and so easily identified. However, the \buone and \agtwo sectors can no longer be distinguished by electron-hole symmetry.

\begin{figure}
    \centering
    \includegraphics[width=0.5\linewidth]{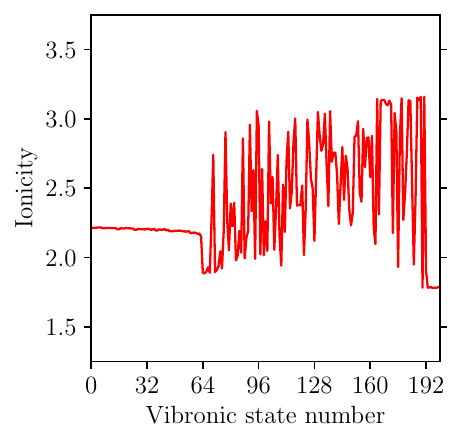}
    \caption{Ionicities of the first $200$ vibronic states for hexatriene with the EHP Hamiltonian and $D=2$. The \agone electronic manifold is well-separated. Entering the \agtwo and \buone electronic manifolds, there are vibronic eigenstates with a range of ionicities due to mixing of the two electronic states by vibrations. }
    \label{fig:qph2_phs0_ionicity}
\end{figure}

Once again we use ionicity to distinguish between the \agtwo and \buone electronic manifolds. Looking beyond the $64^{\text{th}}$ state,~\cref{fig:qph2_phs0_ionicity} shows large fluctuations between high and low ionicity, because the \agtwo and \buone sectors overlap in energy. The high ionicity values, above $3.0$, correspond to vibronic states of primarily \buone electronic character. Low ionicity values, around $2.0$, correspond to vibronic states of primarily \agtwo electronic character. Vibronic eigenstates that lie in between these values have contributions from both electronic states.

\subsection{Extended Hubbard-Peierls Electronic States}
To further confirm ionicity as a good proxy of \bup electronic character and to demonstrate that the higher-energy excited states are well-separated in energy, we present some calculations of the extended Hubbard-Peierls (EHP) Hamiltonian at two different nuclear geometries: The ground-state geometry of the EHP Hamiltonian using \textit{classical} nuclei, in~\cref{tab:hex_states}, and the expectation value of the ground-state geometry of the EHP Hamiltonian using \textit{quantized} nuclei, in~\cref{tab:hex_states_qph_gs}. Because the fully quantum EHP ground state is less dimerized than its classical counterpart, both the \agtwo and \buone have smaller excitation energies. The \agtwo energy decreases more so that the dark-bright energy gap decreases\cite{valentineHigherenergyTripletpairStates2020}. In both cases, the higher-energy excited states, \butwo and $\mathrm{3A_g}$, are well-separated in energy. Moreover, the \butwo state has a much smaller transition dipole moment to the ground state than the \buone state, so it will make a smaller contribution to the initial state.

\begin{table}[ht!]
    \centering
    \begin{tabular}{|c|c|c|c|}
        \hline
        Adiabat & Energy / $\unit{\eV}$ & Ionicity & TDM  \\
        \hline
        \agone & $-23.04$ / $0.00$ & $2.24$ & $0$\\
        \hline
        \buone & $-19.11$ / $3.93$ & $3.15$ & $-1.31$ \\
        \hline
        \agtwo & $-18.82$ / $4.22$ & $1.91$ & $0$  \\
        \hline
        \butwo & $-17.69$ / $5.35$ & $1.91$ & $-0.39$  \\
        \hline
        \agthree & $-16.75$ / $6.29$ & $2.50$ & $0$  \\
        \hline
    \end{tabular}
    \caption{Vertical energies, energy gaps to the ground state, ionicities and transition dipole moments (TDM) of the five lowest-lying eigenstates of the hexatriene EHP Hamiltonian using \textit{classical} nuclei. 
    }
    \label{tab:hex_states}
\end{table}

\begin{table}[ht!]
    \centering
    \begin{tabular}{|c|c|c|c|}
        \hline
        Adiabat & Energy / $\unit{\eV}$ & Ionicity & TDM \\
        \hline
        \agone & $-23.01$ / $0.00$ & $2.23$ & $0$  \\
        \hline
        \buone & $-19.24$ / $3.76$ & $2.76$ & $-1.34$ \\
        \hline
        \agtwo & $-19.14$ / $3.87$ & $1.89$ & $0$ \\
        \hline
        \butwo & $-17.92$ / $5.08$ & $1.85$ & $-0.37$\\
        \hline
        \agthree & $-16.87$ / $6.14$ & $2.54$ & $0$\\
        \hline
    \end{tabular}
    \caption{Vertical energies, energy gaps to the ground state, ionicities and transition dipole moments (TDM) of the five lowest-lying eigenstates of the hexatriene EHP Hamiltonian using \textit{quantized} nuclei. 
    }
    \label{tab:hex_states_qph_gs}
\end{table}

\section{Convergence}\label{apx:convergence}
\begin{figure}
    \centering
    \includegraphics[width=0.5\linewidth]{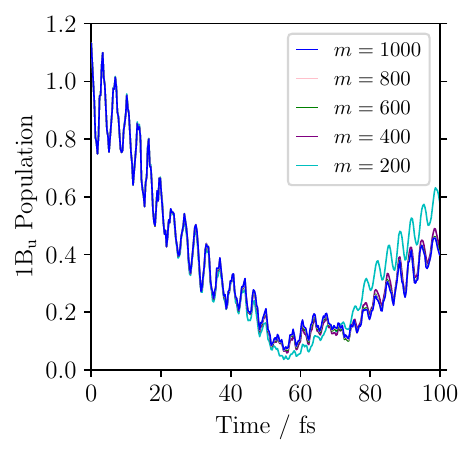}
    \caption{The population of the \buone electronic manifold plotted against time, varying the truncation parameter $m$ of tDMRG. The $m=200$ and $m=400$ results have RMSD greater than $0.01$, relative to the $m=1000$ result. Increasing the truncation cut-off further to $m=600$ and $m=800$ leads to RMSD below $0.01$.}
    \label{fig:m_convergence}
\end{figure}

In the main text, we showed convergence in the number of oscillator levels per site, $D$. tDMRG introduces two further sources of error - the truncation cut-off, $m$, and the time step, $\delta t$.~\cref{fig:m_convergence} shows the \buone population plotted against time for tDMRG simulations with different truncation cut-offs, a time step of $\delta t=0.003~\unit{\fs}$ and $D=6$ oscillator levels per site. Comparing to the $m=1000$ reference calculation, the $m=200$ and $m=400$ have root-mean-squared deviations (RMSD) of $0.085$ and $0.018$, respectively. Both the $m=600$ and $m=800$ simulations have RMSDs below $0.01$, and so we use $m=800$ for our main results.

\begin{figure}
    \centering
    \includegraphics[width=0.5\linewidth]{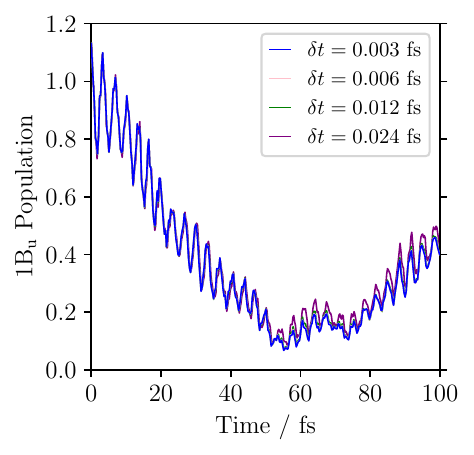}
    \caption{The population of the \buone electronic manifold plotted against time, varying the time step $\delta t$ of tDMRG. The $\delta t=0.006~\unit{\fs}$ and $\delta t=0.012~\unit{\fs}$ calculations have RMSD less than $0.01$, compared to the $\delta t=0.003~\unit{\fs}$ calculation.}
    \label{fig:dt_convergence}
\end{figure}

\cref{fig:dt_convergence} shows the \buone population plotted against time for tDMRG simulations with different time steps, a truncation cut-off of $m=800$ and $D=6$ oscillator levels per site. Comparing to the $\delta t=0.003~\unit{\fs}$ calculation, both the $\delta t=0.006~\unit{\fs}$ and $\delta t=0.012~\unit{\fs}$ calculations have an RMSD below $0.01$. Since reducing the time step in this manner leads to small errors, we use $\delta t=0.003~\unit{\fs}$ for our main results.

\end{document}